\documentclass[12pt]{article}
\usepackage{epsfig}
\usepackage{times}
\newtheorem{theorem}{Theorem}
\newtheorem{lemma}{Lemma}[section]
\newtheorem{remark}{Remark}[section]

\date{}

\begin{document} 

\title{{\LARGE\sf Percolation in the Sherrington-Kirkpatrick
Spin Glass}}
\author{
{\bf J. Machta} 
\\
{\small \tt machta\,@\,physics.umass.edu}\\
{\small \sl Dept. of Physics}\\
{\small \sl University of Massachusetts}\\
{\small \sl Amherst, MA 01003, USA}
\and 
{\bf C. M. Newman}
\\
{\small \tt newman\,@\,cims.nyu.edu}\\
{\small \sl Courant Institute of Mathematical Sciences}\\
{\small \sl New York University}\\
{\small \sl New York, NY 10012, USA}
\and
{\bf D. L. Stein}
\\
{\small \tt daniel.stein\,@\,nyu.edu}\\
{\small \sl Dept.\ of Physics and Courant Institute of
Mathematical Sciences}\\
{\small \sl New York University}\\
{\small \sl New York, NY 10012, USA}
}

\maketitle

\begin{abstract}
We  present extended versions and give
detailed proofs of results concerning percolation 
(using various sets of two-replica bond occupation variables)
in Sherrington-Kirkpatrick
spin glasses (with zero external field) that were first given
in an earlier paper by the same authors. We also explain how
ultrametricity is manifested by the densities of large
percolating clusters.
Our main theorems concern the connection between
these densities and the usual spin overlap
distribution. Their corollaries are that
the ordered spin glass phase is characterized by
a unique percolating cluster of maximal density (normally coexisting
with a second cluster of nonzero but lower density). The proofs
involve comparison inequalities between SK multireplica bond occupation
variables and the independent variables of standard Erd{\H o}s-R{\'e}nyi random
graphs. 

\end{abstract}


\vfill\eject

\small
\renewcommand{\baselinestretch}{1.25}
\normalsize

\section{Introduction}
\label{sec:intro}

In Ising ferromagnets (with no external field), it is well known that the
ordered (broken symmetry) phase manifests itself within the associated
Fortuin-Kasteleyn (FK) random cluster 
representation~\cite{FK72} by the occurrence
of a single positive density percolating cluster (see~\cite{BGJ96}). 
In a recent paper~\cite{MNS07}, we investigated the nature of spin glass
ordering within the FK and other graphical representations and concluded
that the percolation signature of the spin glass phase is the presence
of a single {\it two-replica\/} percolating network of {\it maximal\/}
density, which typically coexists with a second percolating network
of lower density. 
The evidence presented in that paper for this conclusion was 
two-fold: suggestive
numerical results in the case of the three-dimensional Edwards-Anderson (EA)
spin glass~\cite{EA75} and rigorous results for the Sherrington-Kirkpatrick (SK)
spin glass~\cite{SK75}.

In this paper, we expand on those results for the SK model
in several ways. First, we give much more detailed proofs, 
both for two-replica FK (TRFK) percolation and for
the different percolation of ``blue'' bonds in the two-replica graphical
representation studied earlier by 
Chayes, Machta and Redner~\cite{CMR98,CRM98} (CMR). 
Second, we go beyond 
the $\pm J$ SK model (as treated in~\cite{MNS07})
to handle quite general choices of the underlying distribution
$\rho$ for the individual coupling variables, including the usual Gaussian
case. Third, we organize the results (see in particular 
Theorems~\ref{thm:CMR1} and~\ref{thm:TRFK1}) in such a way as
to separate out (see Theorems~\ref{thm:CMR2} and~\ref{thm:TRFK2})
those properties of the overlap distribution 
for the supercritical SK model that are needed to prove related
properties about percolation structure. Such a separation is
called for because many properties of the overlap
distribution that are believed to be valid based on the Parisi
ansatz (see~\cite{MPV87}) for the SK model have not yet been rigorously proved. 
  
Another way we expand on the results of the previous paper
is to present (in Section~\ref{sec:ultrametric}) 
an analyis of the percolation signature of ultrametricity in the
SK model, which is expected to occur, based on the Parisi analysis,
but has not yet been proved rigorously. That is, we describe 
(see Theorems~\ref{thm:ultrametric1}, 
\ref{thm:ultrametric2}~and~\ref{thm:ultrametric3})
how the percolation cluster structure of mutliple
networks of differing densities in the context of
three replicas would exhibit ultrametricity. We note that
as a spinoff of Theorem~\ref{thm:ultrametric3}, we have
(at least in the SK model --- numerical investigations for
the EA model have not yet been done) a third graphical 
percolation signature 
of the spin glass transition 
beyond the two analyzed in our earlier
work --- namely one  involving uniqueness of the maximum
density percolating network (one out of {\it four\/}
clusters) in a three replica 
mixed CMR-FK representation.

In addition to these extensions of our earlier results
and as we note in a remark at
the end of this introductory section, the 
technical machinery we develop in this paper
can be used to obtain
other results for the SK model, 
such as an analysis of large cluster densities
at or near the critical point. Before getting to that, we first
give an outline of the other sections of the paper.

In Section~\ref{sec:results} we describe the SK models
and the CMR and TRFK percolation
occupation variables we will be dealing with throughout.
We then present our main results, starting with 
Theorems~\ref{thm:CMR1} and~\ref{thm:TRFK1} which relate,
in the limit $N\to \infty$,
the densities of the largest percolation clusters 
to the overlap distribution 
for the CMR and TRFK representations respectively. Then, after
stating a known basic result (Theorem~\ref{thm:subcritical})
about the vanishing of the overlap in
the subcritical (and critical) SK model, we present
Theorems~\ref{thm:CMR2} and~\ref{thm:TRFK2},
which give respectively the CMR and TRFK 
percolation signatures of the SK phase transition,
under various assumptions about the SK overlap distribution.
Results relating percolation structure
and ultrametricity in the SK model are presented in
Section~\ref{sec:ultrametric}. 
Then in Section~\ref{sec:proofs}, we present all the proofs.
That section begins with three lemmas that
are the technical heart of the paper and explain why one
can compare, via two-sided stochastic domination inequalties,
SK percolation occupation variables to the independent variables
of Erd{\H o}s-R{\'e}nyi random graphs~\cite{ER60}. 
A key feature of these comparison
results is that they are done only {\it after conditioning on
the values of the spin variables in all the replicas being 
considered\/}. This feature helps explains why the size of the
overlap is crucial --- because it determines the sizes
of the various ``sectors'' of vertices (e.g., those
where the spins in two replicas agree or those where they
disagree) within which the comparisons can be made.

\begin{remark} 
The first part of Theorem~\ref{thm:TRFK2} says
that for $\beta \leq 1$,
the size of the largest TRFK doubly occupied cluster is
${\mbox {o}} (N)$ or equivalently that its density 
$D_1^{{\mbox {TRFK}}} = {\mbox {o}} (1)$ 
as $N \to \infty$.
But the proof (see the TRFK part of Lemma~\ref{lemma2})
combined with known results about
$G(N,p_N)$, the Erd{\H o}s-R{\'e}nyi random graph with $N$ vertices
and independent edge occupation probability $p_N$, implies quite a bit 
more --- that $D_1^{{\mbox {TRFK}}}$ is $ {\mbox {O}} 
(\log{N}/N)$ for $\beta <1$ and $ {\mbox {O}} (N^{2/3}/N)$
for $\beta =1$. Even more is implied in the
critical case. E.g., in the critical scaling window, where 
$\beta = \beta_N = 1 + \lambda / N^{1/3}$, the largest clusters,
of size proportional to $N^{2/3}$, behave exactly like those
occuring in a pair of independent random graphs 
(see, e.g.,~\cite{Bollobas84, Luczak90, JKLP93, Aldous97}) --- i.e.,
as $N \to \infty$, the limiting distribution of 
$(N^{1/3} D_1^{{\mbox {TRFK}}}, N^{1/3} D_2^{{\mbox {TRFK}}}, \dots)$
is the same as that obtained by taking two independent
copies of $G(N/2, 2(\beta_N)^2 /N)$, combining the sizes of the
largest clusters in the two copies, then rank ordering them
and dividing by $N^{2/3}$. One can also show that for $\beta >1$,
the size of the {\it third\/}
largest TRFK cluster behaves like that of the {\it second\/} largest 
cluster in a single copy of the supercritical 
Erd{\H o}s-R{\'e}nyi random graph --- i.e.,
$ {\mbox {O}}(\log{N})$~\cite{ER60}.
But that derivation requires a strengthened version of Lemma~\ref{lemma2}
and further arguments, which will not be presented in this paper.
\end{remark}

\section{Main Results}
\label{sec:results}

Before stating the main results, we specify
the random variables we will be
dealing with. For specificity, we choose
a specific probabilistic coupling
so that even though we deal with two different
graphical representations, and a range of 
inverse temperatures $\beta$,
we define all our random variables for the system of 
positive integer size $N$
on a single probability space. The corresponding probability measure
will be denoted ${\bf P}_N$ (with ${\bf P}$ denoting probability
more generically).

For each $N$, we have three types of random variables: 
real-valued couplings
$\{J_{ij}\}_{1\leq i<j \leq N}$, Ising $\pm 1$-valued spins 
$\{\sigma_i\}_{1\leq i \leq N}$ and $\{\tau_i\}_{1\leq i\leq N}$ for
each of two replicas, and a variety of percolation $\{0,1\}$-valued
bond occupation variables which we will define below. 
These random
variables and their joint distributions depend on both $N$ and
$\beta$ (although we have suppressed that in our notation), 
but to define them, we rely on other sets of
real-valued random variables not depending on $N$ or
$\beta$: $\{K_{ij}\}_{1\leq i<j < \infty}$ and  
$\{U_{ij}^{\ell}\}_{1\leq i<j < \infty}$ for each replica
indexed by $\ell = 1$ or $2$.
(In later sections, we will consider
more than two replicas.) Each of these sets is an i.i.d. family
and the different sets are mutually independent. 
The $U_{ij}^{\ell}$'s are
independent mean one exponentials and will be used to
define the bond occupation variables (conditionally on the
couplings and spins). The $K_{ij}$'s, which determine
the $J_{ij}$'s for given $N$ (and $\beta$) by $J_{ij} = K_{ij}/\sqrt{N}$,
have as their common distribution a probability measure $\rho$
on the real line about which we make the following assumptions:
$\rho$ is even ($d \rho(x) = d \rho(-x)$) 
with no atom at the origin ($\rho (\{0\}) =0$),
variance one ($\int_{-\infty}^{+\infty} x^2 d\rho(x) \, =1$) and 
a finite moment generating function
($\int_{-\infty}^{+\infty} e^{t x} d\rho(x) \, < \, \infty$ 
for all real $t$). The two most common choices are the Gaussian
(where $\rho$ is a mean zero, variance one normal distribution) and the
$\pm J$ (where $\rho$ is 
$(\delta_1 + \delta_{-1})/2$) spin glasses.  

For a given $N$ and $\beta$, we have already defined the couplings
$J_{ij}$. The conditional distribution, given the couplings,
of the spin
variables $\sigma, \tau$ for the two replicas, is that of
an independent sample from the Gibbs distribution;
i.e., 
\begin{equation}
\label{eqn:gibbs}
\mbox{const} \times \exp
\left[   \beta \sum_{1\leq i < j \leq N} J_{ij}
(\sigma_i\sigma_j+\tau_i\tau_j)\right] \ .
\end{equation}

It remains to define the percolation bond occupation variables of interest,
given the couplings and  the spins. Two of these are the FK (random cluster)
variables --- one set for each replica; 
we will denote these ${\bf n}_{ij}^{\ell}$
for $\ell =1$ (corresponding to the first ($\sigma$) replica)
and $\ell =2$ (corresponding to the second ($\tau$) replica). 
These may be constructed as follows.
For a given $i < j$, if the bond $\{i,j\}$ is unsatisfied in the first
replica --- i.e.,
if $J_{ij} \sigma_i \sigma_j < 0$, then set ${\bf n}_{ij}^{1} =0$;
if the bond is satisfied, then set ${\bf n}_{ij}^{1} = 1$ if 
$U_{ij}^{1} \leq 2 \beta |J_{ij}|$ (i.e., with probability
$1 - \exp{(-2 \beta |J_{ij}|)}$) and otherwise set it to zero. 
Define ${\bf n}_{ij}^{2}$ similarly using the second ($\tau$) replica. 
We will be particularly interested in the percolation
properties of the variables ${\bf n}_{ij}= {\bf n}_{ij}^{1} {\bf n}_{ij}^{2}$
that describe {\it doubly FK-occupied \/}bonds. We will use the 
acronym TRFK (for Two Replica FK) to 
denote various quantities built out of
these variables.

There is another two-replica 
graphical representation, introduced by 
Chayes, Machta and Redner~\cite{CMR98, CRM98}
(which we will denote by CMR) that we will also consider. 
This representation in general is described in terms of three
types of bonds which may be thought of as those that are colored blue or red
or else are uncolored. One way of defining the blue bonds, whose
occupation variables we will denote by ${\bf b}_{ij}$, is 
that ${\bf b}_{ij} =1$ if $\{i,j\}$ is satisfied in
{\it both \/} replicas {\it and \/} also
either ${\bf n}_{ij}^{1} =1$ or ${\bf n}_{ij}^{2} =1$ or both
(which occurs with probability $1 - \exp{(-4 \beta |J_{ij}|)}$);
otherwise ${\bf b}_{ij} =0$. We will be interested
in the percolation properties of the blue bonds.
Although we will not be using them
in this paper, we note that a bond $\{i,j\}$ is colored red if and only if
$\sigma_i \sigma_j \tau_i \tau_j = -1$ (or equivalently $\{i,j\}$ is
satisfied in exactly one of the two replicas) {\it and \/} also that
satisfied bond is FK-occupied (which occurs with
probability $1 - \exp{(-2 \beta |J_{ij}|)}$). 

A key role in the theory of spin glasses, and this will also
be the case for their percolation properties, is played by
the Parisi (spin) overlap. For a given $N$ and $\beta$, this overlap
is the random variable, 
\begin{equation}
\label{eqn:overlap}
Q \, = \, Q(N, \beta) \, = 
\, N^{\,-1}\sum_{1\leq i \leq N}\sigma_i \tau_i  \ .
\end{equation}
Closely related to the overlap are the densities 
(i.e., the fractions of sites out of $N$) $D_a=D_a (N,\beta)$ and 
$D_d = D_d (N,\beta)$
of the collections of sites where 
the spins of the two replicas respectively agree and disagree with
each other. Since $Q=D_a - D_d$ and $D_a + D_d =1$, one can
express $D_{\max}= \max\{D_a, D_d\}$ and $D_{\min} = \min \{D_a, D_d\}$
as $D_{\max}=[1+|Q|]/2$ and $D_{\min} = [1-|Q|]/2$. 

It should be clear from our definitions of the various bond
occcupation variables that if one of $i,j$ is in the collection of agree
sites and the other is in the collection of disagree sites, then 
the bond $\{i,j\}$ is satisfied in exactly one of the two replicas
and so $\{i,j\}$ can neither be a TRFK occupied bond nor a CMR blue
bond. So percolation (i.e., occurrence of giant clusters containing
order $N$ of the sites) can only occur separately within the agree
or within the disagree collections of sites. Our results concern when this
happens and its connection with the spin glass phase transition
in the SK model via the overlap random variable $Q$. 

We will first
state two general theorems relating the occurence of  giant clusters 
to the behavior of $Q$ and
then state a number of corollaries. The corollaries depend for
their applicability on results about the nature of $Q$ in the
SK model, some of which have been and some of which have not 
yet been derived rigorously.
The first theorem concerns CMR percolation. We
denote the density of the $k$'th largest CMR blue cluster
by $D_{k}^{\mbox{CMR}}(N,\beta)$.

\begin{theorem}
\label{thm:CMR1}
In the CMR representation, for any $0< \beta < \infty$,
the following three sequences of random variables
tend to zero in probability, i.e., the ${\bf P}_N$-probability
that the absolute value of the random variable is greater than
$\varepsilon$ tends to zero as $N \to \infty$ for any 
$\varepsilon> 0$.
\begin{equation}
\label{eqn:CMR1a}
D_{1}^{\mbox{CMR}}(N,\beta) \, - \, 
[\frac{1+|Q(N,\beta)|}{2}] \, \to \, 0 \, .
\end{equation}
\begin{equation}
\label{eqn:CMR1b}
D_{2}^{\mbox{CMR}}(N,\beta) \, - \,
[\frac{1-|Q(N,\beta)|}{2}] \, \to \, 0 \, .
\end{equation}
\begin{equation}
\label{eqn:CMR1c}
D_{3}^{\mbox{CMR}}(N,\beta) \, \to \, 0 \, .
\end{equation}
\end{theorem}

To state the next theorem, we define $\theta(c)$
for $c \in [0,\infty)$ to be the order parameter for
mean-field percolation --- i.e., the 
asymptotic density (fraction of sites out of $N$
as $N \to \infty$) of the largest
cluster in $G(N,c/N)$, the Erd{\H o}s-R{\'e}nyi random graph
with occupation probability $c/N$ independently for each edge 
in the complete
graph of $N$ sites~\cite{ER60}. It is a standard 
fact~\cite{ER60} that $\theta(c)$ is zero
for $0\leq c \leq 1$ and for $c>1$ is the strictly
positive solution of
\begin{equation}
\theta \, = \, 1 - e^{-c \theta} \,\, .
\end{equation}

\begin{theorem}
\label{thm:TRFK1}
In the TRFK representation, for any $0 <  \beta <  \infty$, the following
limits (in probability) are valid as $N \to \infty$ for 
$D_k^{\mbox{TRFK}}$, the density of the $k$'th largest TRFK doubly occupied
cluster. 
\begin{equation}
\label{eqn:TRFK1a}
D_{1}^{\mbox{TRFK}}(N,\beta) \, - \,
\theta( 2 \beta^2 \frac{1+|Q(N,\beta)|}{2})
[\frac{1+|Q(N,\beta)|}{2}] \, \to \, 0 \, .
\end{equation}
\begin{equation}
\label{eqn:TRFK1b}
D_{2}^{\mbox{TRFK}}(N,\beta) \, - \,
\theta( 2 \beta^2 \frac{1-|Q(N,\beta)|}{2})
[\frac{1-|Q(N,\beta)|}{2}] \, \to \, 0 \, .
\end{equation}
\begin{equation}
\label{eqn:TRFK1c}
D_{3}^{\mbox{TRFK}}(N,\beta) \, \to \, 0 \, .
\end{equation}
\end{theorem}

We now denote by $P(N,\beta)$ the probability distribution
of the overlap $Q(N,\beta)$. This is the Parisi overlap
distribution, averaged over the disorder variables
${\cal K} = \{K_{ij}\}_{1\leq i \leq j < \infty}$. The unaveraged
overlap distribution requires conditioning on ${\cal K}$. So,
for example, we have, for $q \in [-1,1]$,
\begin{equation}
P(N,\beta)([-1,q]) \, = \, {\mbox{Av}} [{\bf P}_N(Q(N,\beta) \leq q|\, 
{\cal K})],
\end{equation}
where ${\mbox{Av}}$ denotes the average over the disorder distribution
of ${\cal K}$.

The quantity $E(Q(N,\beta)^2|\,{\cal K})$ is closely related to 
$\sum_{1\leq i < j \leq N}[E(\sigma_1 \sigma_j|\,{\cal K})]^2$
(see, e.g., \cite{Tal03}, Lemma~$2.2.10$) which in turn is
closely related to the derivative of the finite volume free
energy (see, e.g., \cite{ALR87}, Prop.~$4.1$). It then follows
that $E(Q(N,\beta)^2) \to 0$ as $N \to \infty$ first for
$\beta < 1$ (\cite{ALR87}, Prop.~$2.1$) and then 
(using results of~\cite{GT02,Guerra03,CarHu06} and 
of~\cite{Tal06-1,Tal06-2} --- see also~\cite{Pan07})
also for $\beta =1$. This implies
the following theorem, one of the basic facts in the mathematical
theory of the SK model. 

\begin{theorem}
\label{thm:subcritical}
For $\beta \leq 1$, $Q(N,\beta) \to 0$ (in probability) as $N \to \infty$,
or equivalently $P(N,\beta)\to \delta_0$.
\end{theorem}

The situation regarding rigorous results about the nonvanishing
of $Q(N,\beta)$ as $N \to \infty$ for $\beta > 1$ is less clean.
For example, using results from the references cited just before
Theorem~\ref{thm:subcritical}, it follows that $E(Q(N,\beta)^2)$
has a limit as $N\to \infty$ for all $\beta$, related by a simple
identity to the derivative at that $\beta$ of 
the infinite-volume free energy, ${\cal P} (\beta)$,
given by the Parisi variational formula,
${\cal P} (\beta) = \inf_m \, {\cal P} (m,\beta)$,
where the inf is over distribution functions $m(q)$ with
$q \in [0,1]$ --- see~\cite{Tal06-1}. 
Furthermore, since for $\beta >1$, ${\cal P} (\beta)$ is strictly
below the ``annealed'' free energy~\cite{Co96} (which equals 
${\cal P} (\delta_0,\beta)$, where $\delta_0$ is the distribution
function for the unit point mass at $q=0$), it follows
(see~\cite{Pan07}) from Lipschitz continuity of ${\cal P} (m,\beta)$
in $m$~\cite{Guerra03,Tal06-2} that 
\begin{equation}
\label{supercrit}
\lim_{N\to \infty} E(Q(N,\beta)^2) \, > \, 0 \, {\mbox { for all }} \,
\beta> 1\, .
\end{equation}
However, 
it seems that it
is not yet proved in general that $Q(N,\beta)$ has a 
unique limit (in distribution)
nor very much about the precise nature of any limit.
In order to explain the corollaries
of our main theorems without getting bogged down in these unresolved
questions about the SK model, we will list various properties
which are expected to be valid for (at least some values of) $\beta > 1$
and then use those as assumptions in our corollaries.
Some related comments are given in Remark~\ref{remark2.2} below. 

{\bf Possible Behaviors of the Supercritical Overlap.} For
$\beta > 1$, $P(N,\beta)$ converges as $N\to \infty$
to some $P_\beta$
with the following properties.
\begin{itemize}
\item Property P1: $P_\beta(\{0\})=0$.
\item Property P2: $P_\beta(\{-1,+1\})=0$.
\item Property P3: $P_\beta([-1,-1+(1/\beta^2)]\cup [1-(1/\beta^2),1])=0$.
\end{itemize} 
If one defines $q_{EA}(\beta)$, the Edwards-Anderson order parameter
(for the SK model), to be the supremum of the support of $P_\beta$, and
one assumes that $P_\beta$ has point masses at $\pm q_{EA}(\beta)$,
then Properties P2 and P3 reduce respectively to $q_{EA}(\beta)<1$
and $q_{EA}(\beta)<1-(1/\beta^2)$.
Weaker versions of the three properties that do not require existence
of a limit for $P(N, \beta)$ as $N \to \infty$ are as follows.
\begin{itemize}
\item Property P$1'$: 
\begin{equation}
\label{P1'}
\lim_{\varepsilon \downarrow 0} \limsup_{N \to \infty}
{\bf P}_N(|Q(N,\beta)| < \varepsilon) \, = \, 0.
\end{equation}
\item Property P$2'$:
\begin{equation}
\lim_{\varepsilon \downarrow 0} \limsup_{N \to \infty}
{\bf P}_N(|Q(N,\beta)| > 1-\varepsilon) \, = \, 0.
\end{equation}
\item Property P$3'$
\begin{equation}
\lim_{\varepsilon \downarrow 0} \limsup_{N \to \infty}
{\bf P}_N(|Q(N,\beta)| > 1-(1/\beta^2)-\varepsilon) \, = \, 0.
\end{equation}
\end{itemize}

We will state the next two theorems in a somewhat informal manner
and then provide a more precise meaning in 
Remark~\ref{remark2.1} below.
\begin{theorem}
\label{thm:CMR2}
{\bf (Corollary to Theorem~\ref{thm:CMR1})} In the CMR representation,
for any $0<\beta \leq1$, there are exactly two giant blue clusters,
each of (asymptotic) density $1/2$. For $1 < \beta < \infty$, there
are either one or two giant blue clusters, whose densities add to $1$; there 
is a unique one of (maximum) density in $(1/2,1]$ providing
Property $P1$ (or $P1'$) is valid and there is another one of smaller
density in $(0,1/2)$ providing Property $P2$ (or $P2'$) is valid. 
\end{theorem}

\begin{theorem}
\label{thm:TRFK2}
{\bf (Corollary to Theorem~\ref{thm:TRFK1})} In the TRFK representation,
there are no giant doubly occupied clusters for $\beta \leq 1$. For
$1 < \beta < \infty$, there are either one or two giant doubly
occupied clusters with a unique one of maximum
density providing Property $P1$ (or $P1'$) is valid and 
another one of smaller (but nonzero) density providing Property $P3$
(or $P3'$) is valid.
\end{theorem}

\begin{remark}
\label{remark2.1}
For $0<\beta \leq 1$, Theorem~\ref{thm:CMR2} states
that $(D^{\mbox{CMR}}_1(N,\beta),D^{\mbox{CMR}}_2(N,\beta),
D^{\mbox{CMR}}_3(N,\beta))$  converges (in probability
or equivalently in distribution) to $(1/2,1/2,0)$ while 
Theorem~\ref{thm:TRFK2} states that the corresponding triple of
largest TRFK cluster densities converges to $(0,0,0)$. A precise 
statement of the results for $\beta \in (1,\infty)$ is a bit messier
because it has not been proved that there is a {\it single\/} limit
in distribution of these cluster densities, although since the
densities are all bounded (in $[0,1]$) random variables,
there is compactness with limits along subsequences of $N$'s.
For example, in the CMR case, assuming Properties $P1'$ and $P2'$,
the precise statement is that {\it any\/} limit in distribution
of the triplet of densities is supported on $\{(1/2+a,1/2-a,0):
\, a\in(0,1/2) \}$. Precise statements for the other cases
treated in the two theorems are analogous. 
\end{remark}

\begin{remark}
\label{remark2.2}
Although Property $P1'$ does not seem to have yet been rigorously
proved (for any $\beta >1$), a weaker property does follow 
from~(\ref{supercrit}).
Namely, that 
for all $\beta > 1$,
the limit in~(\ref{P1'}) is strictly
less than one. Weakened versions of portions of Theorems 
\ref{thm:CMR2} and~\ref{thm:TRFK2} for $\beta > 1$ 
follow --- e.g., any limit in distribution of the triplet of densities
in Remark~\ref{remark2.1} must assign strictly positive probability
to $\{(1/2+a,1/2-a,0): \, a\in(0,1/2] \}$.
\end{remark}

\section{Ultrametricity and Percolation}
\label{sec:ultrametric}
In this section, in order to discuss ultrametricity, which
is expected to occur in the supercritical
SK model (see~\cite{MPV87}), we consider three replicas, whose spin variables
are denoted $\{\sigma_i^\ell\}$ for $\ell=1,2,3$. We denote by
${\bf n}_{ij}^\ell$ the FK occupation variables for replica $\ell$
and by ${\bf b}_{ij}^{\ell m}$ the CMR blue bond occupation
variables for the pair of replicas $\ell, m$. Thus
${\bf b}_{ij}^{12}$ corresponds in our previous notation to
${\bf b}_{ij}$. We also denote by $Q^{\ell m}=Q^{\ell m}(N,\beta)$
the overlap defined in~(\ref{eqn:overlap}), but with
$\sigma,\tau$ replaced by $\sigma^\ell,\sigma^m$.

Let us denote by $P^3(N,\beta)$ the distribution of the triple
of overlaps $(Q^{12},Q^{13},Q^{23})$. Ultrametricity concerns the nature of 
the limits as $N\to \infty$ of $P^3(N,\beta)$, as follows,
where we define
\begin{equation}
\label{eqn:ultrametric1}
{\bf R}_{\mbox{ultra}}^3 \, = \, \{(x,y,z):\, |x|=|y|\leq|z|\, {\mbox{or}}
\, |x|=|z|\leq|y| \, {\mbox{or}}\, |y|=|z|\leq|x|\} \, .
\end{equation}

{\bf Possible Ultrametric Behaviors of the Supercritical Overlap.} For
$\beta>1$, $P^3(N,\beta)$ converges to some $P_\beta ^3$ as $N \to \infty$
with
\begin{itemize}
\item Property P4: $P_\beta ^3({\bf R}_{\mbox{ultra}}^3)=1$.
\end{itemize}
We will generally replace this property by a weakened version, 
P4$'$, in which it is not assumed that there is a single
limit $P_\beta ^3$ as $N \to \infty$ 
but rather the same property is assumed for
every subsequence limit. 
There is another property that
simplifies various of our statements about
how ultrametricity is manifested in the sizes of various percolation
clusters. This property, which, like ultrametricity, is expected to be valid in
the supercritical SK model (see~\cite{CGGPV07}, where this property
is discussed and also numerically tested in the three-dimensional
EA model) is the following.
\begin{itemize}
\item Property P5: 
$P_\beta ^3(\{(x,y,z):\, xyz\geq0 \})=1$.
\end{itemize}
Again we will use a weaker version P5$'$ in which it is not
assumed that there is a single limit $P_\beta ^3$ as $N\to \infty$. 

One formulation of ultrametricity using percolation clusters
is the next theorem,
an immediate corollary of Theorem~\ref{thm:CMR1},
in which we denote by $D_j^{\ell m}=D_j^{\ell m}(N,\beta)$ the density
of sites in ${\cal {C}}_j^{\ell m}$, the $j$'th largest cluster
formed by the bonds $\{i,j\}$ with ${\bf b}_{ij}^{\ell m} =1$ (i.e.,
the $j$'th largest CMR blue cluster for the pair $\{\ell,m\}$ of replicas).
Note that $D_j^{12}$ coincides in our previous notation
with $D_j^{\mbox{CMR}}$.
\begin{theorem}
\label{thm:ultrametric1}
{\bf (Corollary to Theorem~\ref{thm:CMR1})} For $1< \beta < \infty$,
assuming Property P4$'$, any subsequence limit 
in distrbution as $N \to \infty$ of
the triple $(D_1^{12}-D_2^{12},D_1^{13}-D_2^{13},D_1^{23}-D_2^{23})$
is supported on ${\bf R}_{\mbox{ultra}}^3$. 
\end{theorem}

In our next two theorems, instead of looking at {\it differences\/} of 
densities, we express ultrametricity directly in terms of densities themselves.
This is perhaps more interesting because rather than having three
density differences, there will be {\it four\/} densities. We begin
in Theorem~\ref{thm:ultrametric2}
with a fully CMR point of view with
four natural non-empty {\it intersections\/}
of CMR blue clusters.
Then Theorem~\ref{thm:ultrametric3} 
mixes CMR and FK
occupation variables to yield (four) other natural clusters.

There are a number of ways in which the four
sets of sites in our fully CMR perspective can 
be defined, which turn out to be equivalent (for
large $N$). 
One definition is as follows. For $\alpha,\alpha'$ each taken
to be either the letter $a$ (for agree) or the letter $d$ 
(for disagree), define $\Lambda_{\alpha \alpha'}(N,\beta)$ to
be the set of sites $i \in\{1,\dots,N\}$ where $\sigma_i^1$
agrees (for $\alpha = a$) or disagrees (for $\alpha = d$) with
$\sigma_i^2$ and $\sigma_i^1$ agrees (for $\alpha' = a$) or disagrees (for 
$\alpha' = d$) with $\sigma_i^3$; also denote by $D_{\alpha \alpha'}(N,\beta)$
the density of sites (i.e., the fraction of $N$)
in $\Lambda_{\alpha \alpha'}(N,\beta)$. 
Then denote by ${\cal C}_{\alpha \alpha'}^{\ell m}$ the largest
cluster (thought of as the collection of its sites)
formed within $\Lambda_{\alpha \alpha'}$ by the
${\bf b}^{\ell m}=1$ blue bonds. Finally, define 
${\cal C}_{\alpha \alpha'} = {\cal C}_{\alpha \alpha'}^{12}
\cap {\cal C}_{\alpha \alpha'}^{13}$, 
$D_{\alpha \alpha'}^{\mbox{CMR}}$ to be the density of
sites (fraction of $N$) in ${\cal C}_{\alpha,\alpha'}$ and
${\hat D}^{\mbox{CMR}}(N,\beta)$ to be the vector of
four densities $(D_{aa}^{\mbox{CMR}},D_{ad}^{\mbox{CMR}},
D_{da}^{\mbox{CMR}},D_{dd}^{\mbox{CMR}})$. To state
the next theorem, let 
${\bf R}^{4,+}$ denote $\{(x_1,x_2,x_3,x_4):
{\mbox {each}}\, x_i \geq 0\}$ and define 
\begin{equation}
{\bf R}^{4,+}_{\mbox{ultra}}=\{(x_1,x_2,x_3,x_4)\in{\bf R}^{4,+}:
\, x_{(1)}>x_{(2)}\geq x_{(3)}=x_{(4)} \},
\end{equation}
where $x_{(1)},x_{(2)},x_{(3)},x_{(4)}$ are the rank ordered
values of $x_1,x_2,x_3,x_4$.
\begin{theorem}
\label{thm:ultrametric2} For $0< \beta \leq 1$,
${\hat D}^{\mbox{CMR}}(N,\beta)$ $ \to (1/4,1/4,1/4,1/4)$
(in probability) as $N\to\infty$. For $1 < \beta < \infty$
and assuming Properties P1$'$, P2$'$, P4$'$,P5$'$,
any limit in distribution of ${\hat D}^{\mbox{CMR}}(N,\beta)$
is supported on 
${\bf R}_{\mbox{ultra}}^{4,+}\,$ .
\end{theorem}

\begin{remark}
\label{remark3.1}
The equilateral triangle case where
$|Q^{12}|=|Q^{13}|=|Q^{23}|=q_u$ corresponds to
$x_{(1)}=(1+3q_u)/4$ and $x_{(2)}=x_{(3)}=x_{(4)}=(1-q_u)/4$.
The alternative isosceles triangle case with, say, 
$|Q^{12}|=q_u > |Q^{13}|=|Q^{23}|=q_\ell$ corresponds to
$x_{(1)}=(1+q_u+2q_\ell)/4$, $x_{(2)}=(1+q_u-2q_\ell)/4$
and $x_{(3)}=x_{(4)}=(1-q_u)/4$.
\end{remark}

In the next theorem, we consider ${\cal C}_{\alpha\alpha'}^{*}$
defined as the largest cluster within $\Lambda_{\alpha\alpha'}(N,\beta)$
formed by bonds $\{i,j\}$ with ${\bf b}_{ij}^{12} {\bf n}_{ij}^{3}=1$ --- i.e.,
bonds that are simultaneously CMR blue for the first 
two replicas and FK-occupied 
for the third replica. Let ${\hat D}^{*}(N,\beta)$ denote
the corresponding vector of four densities. As in the previous theorem,
we note that there are alternative, but
equivalent for large $N$, definitions of the clusters
and densities (e.g., as the four largest clusters formed by bonds
$\{i,j\}$ with ${\bf b}_{ij}^{12} {\bf n}_{ij}^{3}=1$ 
in all of $\{1,\dots,N\}$ without a priori restriction to
$\Lambda_{\alpha\alpha'}(N,\beta)$).

\begin{theorem}
\label{thm:ultrametric3} For $0< \beta \leq 1$, 
${\hat D}^{*}(N,\beta) \to (0,0,0,0)$
(in probability) as $N \to \infty$. For $1< \beta < \infty$ and
assuming Properties P1$'$, P3$'$, P4$'$, P5$'$, any limit
in distribution of ${\hat D}^{*}(N,\beta)$
is supported on ${\bf R}_{\mbox{ultra}}^{4,+}\,$ .
\end{theorem}

\begin{remark}
\label{remark3.2}
Here the limiting densities are of the form
$x_{(j)}=\theta(4 \beta^2 d_{(j)}) \, d_{(j)}$ 
where $d_{(1)}=(1+q_u+2q_\ell)/4$,
$d_{(2)}=(1+q_u-2q_\ell)/4$ and $d_{(3)}=d_{(4)}=(1-q_u)/4$,
with $q_\ell=q_u$ for the equilateral triangle case.
\end{remark}

\section{Proofs}
\label{sec:proofs}

Before giving the proofs of our main results, we present several key lemmas
which are the technical heart of our proofs. We will use the notation
$>>$ and $<<$ to denote stochastic domination (in the FKG sense) for either
families of random variables or their distributions. E.g., for
the $m$-tuples $X=(X_1,\dots,X_m)$ and $Y=(Y_1,\dots,Y_m)$ we write
$X<<Y$ or $Y>>X$ to mean that ${\bf E}(h(X_1,\dots,X_m)) \leq
{\bf E}(h(Y_1,\dots,Y_m))$ for every coordinatewise increasing function
$h$ (for which the two expectations exist). 

Our key lemmas
concern stochastic domination inequalities in
the $k$-replica setting comparing conditional distributions
of the couplings $\{K_{ij}\}$ or related bond
occupation variables, when the spins
$\sigma^1,\dots,\sigma^k$ are fixed, to product
measures. These allow us to approximate percolation
variables in the SK model by the {\it independent\/} variables
of Erd{\H o}s-R{\'e}nyi random graphs when $N\to\infty$. 
Given probability measures $\nu_{ij}$ on ${\bf R}$
for $1\leq i<j \leq N$, we denote
by ${\mbox{Prod}}_N(\{\nu_{ij}\})$ the corresponding product measure on 
${\bf R}^{N(N-1)/2}$. We also will denote by $\rho[\gamma]$ the
probability measure defined by $d\rho[\gamma](x) = e^{\gamma x} d\rho(x)/
\int_{-\infty}^{+\infty}e^{\gamma x'} d\rho(x')$.

\begin{lemma}
\label{lemma1}
Fix $N,\beta,k$ and let ${\tilde \mu}_{N,\beta}^k$ denote
the conditional distribution, given the spins 
$\sigma^1=(\sigma_i^1:1 \leq i \leq N),\dots,\sigma^k$,
of $\{K'_{ij} \equiv \varepsilon_{i,j}K_{ij}\}_{1 \leq i < j \leq N}$
(where the $\varepsilon_{i,j}$'s are any given $\pm1$ values). Then
\begin{equation}
\label{eqn:FKG1}
{\mbox{Prod}}_N(\{\rho[\gamma_{ij}^{k,-}] \}) \, << \,
{\tilde \mu}_{N,\beta}^k  \, << \,
{\mbox{Prod}}_N(\{\rho[\gamma_{ij}^{k,+}] \}) \, ,
\end{equation}
where
\begin{equation}
\gamma_{ij}^{k,\pm} = \, \frac{\beta}{\sqrt{N}} 
[\varepsilon_{ij}(\sigma_i^1 \sigma_j^1+ \dots + \sigma_i^k \sigma_j^k)
\pm k] \, .
\end{equation}
\end{lemma}

\noindent {\bf Proof.} 
Define the partition function 
\begin{equation}
\label{eqn:partition}
Z_{N,\beta}=Z_{N,\beta}(\{K'_{ij}\}) = \,
\sum_\sigma \exp({\frac{\beta}{\sqrt{N}} \sum_{1\leq i < j \leq N}
(\varepsilon_{ij} \sigma_i \sigma_j)K'_{ij}}) \, ,
\end{equation}
where $\sum_\sigma$ denotes the sum over all $2^N$ choices of 
$\sigma_i=\pm 1$ for $1 \leq i \leq N$. 
Thus the normalization constant in Equation~(\ref{eqn:gibbs}) for two
replicas is $(Z_{N,\beta})^{-2}$ and the 
$k$-replica marginal distribution for $\{K'_{ij}\}$ is as follows,
where ${\mbox {Prod}}_N(\{\rho\})$ denotes ${\mbox {Prod}}_N(\{\nu_{ij}\})$
with $\nu_{ij} \equiv \rho$, $\phi_{ij}^k = \varepsilon_{ij}
(\sigma_i^1 \sigma_j^1 + \dots + \sigma_i^k \sigma_j^k)$
and $C_{N,\beta}$ is a normalization constant depending on 
the $\phi_{ij}^k$'s but not on the $K'_{ij}$'s.
\begin{equation}
d{\tilde \mu}_{N,\beta}^k = \, C_{N,\beta} (Z_{N,\beta}(\{K'_{ij}\}))^{-k}
\exp({\frac{\beta}{\sqrt{N}} \sum_{1\leq i < j \leq N}
\phi_{ij}^k K'_{ij}}) \, {\mbox {Prod}}_N(\{\rho\}) \, .
\end{equation}

It is a standard fact about stochastic domination that if ${\tilde \mu}$
is a product probability 
measure on ${\bf R}^m$ and $g$ is an increasing --- i.e., 
coordinatewise nondecreasing --- (respectively, decreasing)
function on ${\bf R}^m$ 
(with $\int g d {\tilde \mu} \, < \infty$), then 
the probability measure  ${\tilde \mu}_g$ defined as
$d{\tilde \mu}_g(x_1,\dots,x_m) = g(x_1,\dots,x_m) d{\tilde \mu}/
\int g d {\tilde \mu}$ satisfies ${\tilde \mu}_g >> {\tilde \mu}$
(respectively, ${\tilde \mu}_g  << {\tilde \mu}$). This follows
from the fact that product measures satisfy the FKG 
inequalities --- i.e., for $f$ and $g$ increasing
$\int f g d {\tilde \mu} \geq \int f d {\tilde \mu} \, \int g d {\tilde \mu}$.

On the other hand, since each $\varepsilon_{ij} \sigma_i \sigma_j = \pm 1$,
it follows from~(\ref{eqn:partition}) that 
\begin{equation}
Z_{N,\beta}^{\pm} \equiv Z_{N,\beta} 
e^{\pm(\beta/\sqrt{N}) \sum K'_{ij}}
= \, \sum_\sigma e^{(\beta/\sqrt{N}) \sum \psi_{ij}^{\pm} (\sigma) K'_{ij}}
\end{equation}
with each $\psi_{ij}^{\pm} = 0$ or~$\pm2$ 
and hence each $\psi_{ij}^{+} \geq 0$ (respectively,
each $\psi_{ij}^{-} \leq 0$). Thus, as a function of the $K'_{ij}$'s,
$Z_{N,\beta}^{+}$ is increasing and $(Z_{N,\beta}^{+})^{-k}$ is decreasing
while $Z_{N,\beta}^{-}$ is decreasing and $(Z_{N,\beta}^{-})^{-k}$ is 
increasing. Combining this with the previous discussion about 
stochastic domination and product measures, we see that
\begin{equation}
\label{eqn:FKG2}
C_{N,\beta}^{-} e^{-(\beta/\sqrt{N}) \sum(-kK'_{ij}+\phi_{ij}^k K'_{ij})}
\, {\mbox {Prod}}_N(\{\rho\}) \, << \, {\tilde \mu}_{N,\beta}^k << \, 
C_{N,\beta}^{+} e^{-(\beta/\sqrt{N}) \sum(+kK'_{ij}+\phi_{ij}^k K'_{ij})}
\, {\mbox {Prod}}_N(\{\rho\}) \, ,
\end{equation}
which is just Equation~(\ref{eqn:FKG1}) since the $C_{N,\beta}^{\pm}$
are normalization constants. This completes the 
proof of Lemma~\ref{lemma1}.

The next two lemmas give stochastic domination inequalities for
three different sets of occupation variables in terms of independent
Bernoulli percolation variables. The first lemma covers the
cases of CMR and TRFK variables involving two replicas and the
second lemma deals with mixed CMR-FK variables involving three replicas. 
The parameters of the independent Bernoulli variables used for upper and lower
bounds are denoted $p_{N,\beta}^{*,\natural}$, where
* denotes CMR or TRFK or $3$ (for the mixed CMR-FK case) and $\natural$ denotes
$u$ (for upper bound) or $\ell$ (for lower bound) and are as follows.
\begin{equation}
\label{eqn:param1}
p_{N,\beta} ^{*,u} = \, 
\int_0 ^\infty g_{N,\beta}^{*} (x) d\rho_{N,\beta}^{*} (x) \, ,
\end{equation}
\begin{equation}
\label{eqn:param2}
p_{N,\beta}^{*,\ell} = \, 
\int_0^\infty g_{N,\beta}^{*} (x) d\rho(x) \, ,
\end{equation}
where
\begin{equation}
\label{eqn:param3}
g_{N,\beta}^{{\mbox {CMR}}} (x) = 1-e^{-4(\beta / \sqrt{N})x} \, ,
\end{equation}
\begin{equation}
\label{eqn:param4}
g_{N,\beta}^{{\mbox {TRFK}}} (x) = (1-e^{-2(\beta/ \sqrt{N})x})^2 \, ,
\end{equation}
\begin{equation}
\label{eqn:param5}
g_{N,\beta}^{3} (x) = (1-e^{-4(\beta/ \sqrt{N})x}) \,
(1-e^{-2(\beta/ \sqrt{N})x}),
\end{equation}
and
\begin{equation}
\label{eqn:param6}
\rho_{N,\beta}^{{\mbox {CMR}}} = \, 
\rho_{N,\beta}^{{\mbox {TRFK}}} = \,
\rho[4\beta/ \sqrt{N}] \, ,
\end{equation}
\begin{equation}
\label{eqn:param7}
\rho_{N,\beta}^{3}= \, \rho[6\beta/ \sqrt{N}] \, .
\end{equation}

\begin{lemma}
\label{lemma2}
Fix $N, \beta, \sigma^1,\sigma^2$ and consider the conditional distributions
${\hat {\mu}}_{N,\beta}^{{\mbox {CMR}}}$
of $\{{\bf b}_{i,j} \equiv {\bf b}_{i,j}^{12}\}_{1\leq i < j \leq N}$ and
${\hat {\mu}}_{N,\beta}^{{\mbox {TRFK}}}$ of 
$\{{\bf n}_{i,j} \equiv {\bf n}_{i,j}^1 {\bf n}_{i,j}^2\}_{1\leq i < j \leq N}$.
Then
\begin{equation}
\label{eqn:FKG3}
{\mbox {Prod}}_N (p_{i,j}^{*,\ell}\delta_1 + (1-p_{i,j}^{*,\ell})\delta_0)
\, << \, {\hat {\mu}}_{N,\beta}^{*} \, << \,
{\mbox {Prod}}_N (p_{i,j}^{*,u}\delta_1 + (1-p_{i,j}^{*,u})\delta_0) \, ,
\end{equation}
where $p_{ij}^{*,\natural} = p_{N,\beta}^{*,\natural}$ if $i,j$ are
either both in $\Lambda_a$ or both in $\Lambda_d$ and $p_{ij}^{*,\natural} =0$
otherwise.
The asymptotic behavior as $N \to \infty$ of the parameters appearing
in these inequalities is
\begin{equation}
p_{N,\beta}^{{\mbox {CMR}},\natural} = \frac{2 \beta}{\sqrt{N}}
(\int_{-\infty}^{+\infty} |x| d \rho(x)) 
(1 + {\mbox {O}}(\frac{1}{\sqrt{N}})) \, ,
\end{equation}
\begin{equation}
p_{N,\beta}^{{\mbox {TRFK}},\natural} = \frac{2 \beta^2}{N}
(1 + {\mbox {O}}(\frac{1}{\sqrt{N}})) \, .
\end{equation}
\end{lemma}

\noindent{\bf Proof.}
We begin with some considerations for the general case of
$k$ replicas and arbitrary $\varepsilon_{ij}$ before 
specializing to what is used in this lemma. The case $k=3$
will be used for the next lemma. Let $({\tilde K},{\tilde U})_N^k$
$\equiv ({\tilde K}_{ij}, {\tilde U}_{ij}^1, \dots, {\tilde U}_{ij}^k)_N$
denote random variables (with $1\leq i < j \leq N$)
whose joint distribution is the conditional
distribution of $(K'_{ij},U_{ij}^1, \dots,U_{ij}^k)_N$ given
$\sigma^1,\dots,\sigma^k$. Note that the ${\tilde U}_{ij}^\ell$'s
are independent of the ${\tilde K}_{ij}$'s and 
(like the $U_{ij}^\ell$'s) are independent mean one exponential
random variables. Let $(K^{\mbox{ind}}[\{\gamma_{ij}\}],
U^{\mbox{ind}})_N^k$ $\equiv (K^{\mbox{ind}}_{ij}[\gamma_{ij}],
U^{{\mbox{ind}},1}_{ij},\dots, U^{{\mbox{ind}},k}_{ij})_N$ denote
mutually
independent random variables with $K^{\mbox{ind}}_{ij}[\gamma_{ij}]$
distributed by $\rho[\gamma_{ij}]$ and each 
$U^{{\mbox{ind}},\ell}_{ij}$ a mean one exponential random variable.

Lemma~\ref{lemma1} says that 
\begin{equation}
\label{eqn:FKG4}
(K^{\mbox{ind}}[\{\gamma_{ij}^{k,-}\}])_N^k << \, ({\tilde K})_N^k
<< \, (K^{\mbox{ind}}[\{\gamma_{ij}^{k,+}\}])_N^k
\end{equation}
and it immediately follows that
\begin{equation}
\label{eqn:FKG5}
(K^{\mbox{ind}}[\{\gamma_{ij}^{k,-}\}],-U^{\mbox{ind}})_N^k << \, 
({\tilde K},-{\tilde U})_N^k
<< \, (K^{\mbox{ind}}[\{\gamma_{ij}^{k,+}\}],-U^{\mbox{ind}})_N^k \, .
\end{equation}

We now take $k=2$, choose $\varepsilon_{ij} =\sigma^1_i \sigma^1_j$ and
note that ${\bf n}_{ij} \equiv {\bf n}_{ij}^1 {\bf n}_{ij}^2
=0 = {\bf b}_{ij}$ {\it unless\/} $i,j$ are either 
both in $\Lambda_a$ or both in $\Lambda_d$, in which case
we have 
$\sigma^1_i \sigma^1_j = \sigma^2_i \sigma^2_j$,
$\gamma_{ij}^{2,+} = 4\beta/ \sqrt{N}$ and $\gamma_{ij}^{2,-}=0$.
Furthermore ${\bf n}_{ij}^\ell$ is the indicator of the event
that $2 \beta K'_{ij} + (-U_{ij}^\ell) \geq 0$ while
${\bf b}_{ij} \equiv {\bf b}_{ij}^{12}$ is the indicator of the event
that ${\bf n}_{ij}^1 + {\bf n}_{ij}^2 \geq 1$ and $ K'_{ij} \geq0$,
so that these occupation variables (for such $i,j$) are
increasing functions of $(K',-U)_N^2$.
Let us now define ${\tilde {\bf n}}_{ij}, {\tilde {\bf b}}_{ij}$
and ${\bf n}^{{\mbox {ind}},\pm}_{ij}, {\bf b}^{{\mbox {ind}},\pm}_{ij}$
as the same increasing functions,
respectively, of $({\tilde K},-{\tilde U})_N^2$ and
$(K^{\mbox{ind}}[\{\gamma_{ij}^{k,\pm}\}],
-U^{\mbox{ind}})_N^2$ providing $i,j$ 
are either both in $\Lambda_a$ or both in $\Lambda_d$,
and otherwise set these occupation variables to zero. 
Then, as a consequence of~(\ref{eqn:FKG5}),
we have
\begin{equation}
\label{eqn:FKG6}
({\bf n}^{{\mbox {ind}},-})_N^2 << \, ({\tilde {\bf n}})_N^2 << \,
({\bf n}^{{\mbox {ind}},+})_N^2 \, ,
\end{equation}
and
\begin{equation}
\label{eqn:FKG7}
({\bf b}^{{\mbox {ind}},-})_N^2 << \, ({\tilde {\bf b}})_N^2 << \,
({\bf b}^{{\mbox {ind}},+})_N^2 \, .
\end{equation}

To complete the proof of Lemma~\ref{lemma2}, it remains to
obtain the claimed formulas for the various Bernoulli occupation
parameters. E.g., for the case $* = {\mbox {TRFK}}$ and
$\natural = u$, we have for $i,j$ 
either both in $\Lambda_a$ or both in $\Lambda_d \,$,
\begin{eqnarray}
p_{N,\beta}^{{\mbox {TRFK}},u} & = &
{\mbox {\bf P}}({\bf n}^{{\mbox {ind}},+}_{ij} = \, 1)
\nonumber \\
& = &
{\mbox {\bf P}} (2 \beta K^{\mbox {ind}}_{ij}\gamma^{2,+}_{ij}
/\sqrt{N} \, \geq U^{{\mbox {ind}},1}_{ij}, U^{{\mbox {ind}},2}_{ij})
\nonumber \\
& = & \int_0^\infty(1-e^{-2(\beta/ \sqrt{N)}x})^2
d\rho[4 \beta / \sqrt{N}](x) \, ,
\end{eqnarray}
as given by Equations~(\ref{eqn:param1}), (\ref{eqn:param4})
and~(\ref{eqn:param6}). We leave the checking of the other
three cases (for $k=2$) and the straightforward
derivation of the asymptotic behavior
as $N \to \infty$ for all the parameters to the reader.
This completes the proof of Lemma~\ref{lemma2}.

\begin{lemma}
\label{lemma3}
Fix $N, \beta, \sigma^1,\sigma^2,\sigma^3$ 
and consider the conditional distribution
${\hat {\mu}}_{N,\beta}^3$
of $\{{\bf b}_{i,j} {\bf n}_{i,j}^3\}_{1\leq i < j \leq N}$.
Then equation~(\ref{eqn:FKG3}) with $*=3$ remains valid, where
$p_{ij}^{3,\natural} = p_{N,\beta}^{3,\natural}$ if $i,j$ are
either both in $\Lambda_{aa}$ or both in $\Lambda_{ad}$ or
both in $\Lambda_{da}$ or both in $\Lambda_{dd}$ and 
$p_{ij}^{3,\natural} =0$ otherwise.
The asymptotic behavior as $N \to \infty$ of $p_{N,\beta}^{3,\natural}$
is
\begin{equation}
p_{N,\beta}^{3,\natural} = \frac{4 \beta^2}{N}
(1 + {\mbox {O}}(\frac{1}{\sqrt{N}})) \, .
\end{equation}
\end{lemma}

\noindent {\bf Proof.}
As in the proof of Lemma~\ref{lemma2}, we apply Lemma~\ref{lemma1}
in the guise of~(\ref{eqn:FKG5}) with 
$\varepsilon_{ij} =\sigma^1_i \sigma^1_j$ again, but this
time with $k=3$. Now ${\bf b}_{ij} {\bf n}_{ij}^3 = 
{\bf b}_{ij}^{12} {\bf n}_{ij}^3 =0$ {\it unless\/} $i,j$ are either
both in $\Lambda_{aa}$ or both in $\Lambda_{ad}$ 
or both in $\Lambda_{da}$ or both in $\Lambda_{dd}$,
in which case
we have
$\sigma^1_i \sigma^1_j = \sigma^2_i \sigma^2_j
=\sigma^3_i \sigma^3_j$, 
$\gamma_{ij}^{3,+} = 6\beta/ \sqrt{N}$, $\gamma_{ij}^{3,-}=0$,
and ${\bf b}_{ij}^{12} {\bf n}_{ij}^3$ is an increasing function
of $(K',-U)_N^3$.
The remainder of the proof, which closely mimics that of 
Lemma~\ref{lemma2}, is straightforward.

\noindent {\bf {Proof of Theorem~\ref{thm:CMR1}.}}
We denote by $D_{\alpha,j}^{\mbox {CMR}} =
D_{\alpha,j}^{\mbox {CMR}} (N, \beta)$ for $\alpha = a$
or $d$ the density (fraction of $N$) of the $j$'th largest 
CMR cluster in $\Lambda_\alpha$ and by $S_j^{\mbox {RG}} =
S_j^{\mbox {RG}}(N, p_N)$ the number of sites in the largest
cluster of the random graph $G(N,p_N)$. Recall that 
$D_\alpha= D_\alpha(N, \beta)$ denotes the density (fraction of $N$) of
$\Lambda_\alpha$ and that $\max\{D_a, D_d\} = [1+|Q|]/2$,
$\min \{D_a, D_d\}  = [1-|Q|]/2$. Then Lemma~\ref{lemma2}
implies that, conditional on $\sigma^1,\sigma^2$, 
\begin{equation}
\label{eqn:FKG8}
N\, D_{\alpha,1}^{\mbox {CMR}} (N, \beta) \, >> \,
S_1^{\mbox {RG}}(N \,D_\alpha(N, \beta), (2\beta c_\rho/ \sqrt{N}) 
+ {\mbox {O}} (1/N)) 
\end{equation}
with $c_\rho >0$. By separating the cases of small  and not
so small $D_\alpha(N, \beta)$ and using the above stochastic
domination combined with the facts that 
$m^{-1}S_1^{\mbox {RG}}(m,p_m) \to \theta(1) =1$ (in probability)
when $m, mp_m \to \infty$, and that $D_{\alpha,1}^{\mbox {CMR}} 
\leq D_\alpha$, it directly follows that 
$D_\alpha - D_{\alpha,1}^{\mbox {CMR}} \to 0$
and hence also that $D_{\alpha,2}^{\mbox {CMR}} \to 0$
(in probability). This then directly yields~(\ref{eqn:CMR1a}),
(\ref{eqn:CMR1b}) and (\ref{eqn:CMR1c}) and completes the proof
of Theorem~\ref{thm:CMR1}.

\noindent {\bf {Proof of Theorem~\ref{thm:TRFK1}.}}
The proof mimics that of Theorem~\ref{thm:CMR1} except that
(\ref{eqn:FKG8}) is replaced by
\begin{equation}
\label{eqn:FKG9}
S_1^{\mbox {RG}}(N \,D_\alpha, \frac{2 \beta^2}{N}
+ {\mbox {O}} (N^{-3/2})) \, >> \, N\, D_{\alpha,1}^{\mbox {TRFK}} 
\, >> \, 
S_1^{\mbox {RG}}(N \,D_\alpha, \frac{2 \beta^2}{N}
+ {\mbox {O}} (N^{-3/2})) \, ,
\end{equation}
where the two terms correcting $2 \beta^2/N$, while
different, are both of
order ${\mbox {O}} (N^{-3/2})$.
We then use the fact that
$m^{-1}S_1^{\mbox {RG}}(m,p_m) - \theta(mp_m) \to 0$ when
$m \to \infty$ to conclude that 
$D_{\alpha,1}^{\mbox {TRFK}} (N, \beta) - \theta(2 \beta^2 
D_\alpha(N, \beta)) D_\alpha(N, \beta))
\to 0$ which yields~(\ref{eqn:TRFK1a})
and~(\ref{eqn:TRFK1b}). To obtain~(\ref{eqn:TRFK1c}), 
we need to
show that $D_{\alpha,2}^{\mbox {TRFK}} \to 0$.
But this follows
from what we have already showed, from the analogue of~(\ref{eqn:FKG9})
with $D_{\alpha,1}^{\mbox {TRFK}}$ replaced by
$D_{\alpha,1}^{\mbox {TRFK}} + D_{\alpha,2}^{\mbox {TRFK}}$
and $S_1^{\mbox {RG}}$ replaced by $S_1^{\mbox {RG}} + S_2^{\mbox {RG}}$,
and by the fact that $m^{-1}S_2^{\mbox {RG}}(m,p_m) \to0$ in
probability when $m \to \infty$. This completes the proof
of Theorem~\ref{thm:TRFK1}.

\noindent {\bf {Proof of Theorem~\ref{thm:CMR2}.}}
Theorem~\ref{thm:CMR1} implies that any limit in distribution of
$(D_1^{\mbox {CMR}}, D_2^{\mbox {CMR}},D_3^{\mbox {CMR}})$ coincides
with some limit in distribution of $(\frac{1+|Q|}{2},\frac{1-|Q|}{2},0)$.
The claims of the theorem for $0<\beta \leq 1$ then follow from
Theorem~\ref{thm:subcritical} and for $1<\beta < \infty$ from that
theorem and Properties P1$'$ and P2$'$.

\noindent {\bf {Proof of Theorem~\ref{thm:TRFK2}.}}
Theorem~\ref{thm:TRFK1} implies that any limit in distribution of
$(D_1^{\mbox {TRFK}}, D_2^{\mbox {TRFK}},D_3^{\mbox {TRFK}})$
coincides
with some limit in distribution of 
$(\theta(\beta^2 (1+|Q|)) (\frac{1+|Q|}{2}),
\theta(\beta^2 (1-|Q|)) (\frac{1-|Q|}{2}),0)$.
The claims of the theorem for $0<\beta \leq 1$ then follow from
Theorem~\ref{thm:subcritical} and the fact that
$\theta(c)=0$ for $c \leq 1$ while the claims
for $1<\beta < \infty$ follow from Theorem~\ref{thm:subcritical}
and Properties P1$'$ and P3$'$. Note that P3$'$ is
relevant because $\theta(\beta^2 (1-|Q|)) >0$ if and only if
$\beta^2 (1-|Q|)>1$ or equivalently $|Q| > 1 - (1/ \beta^2)$.

\noindent {\bf {Proof of Theorem~\ref{thm:ultrametric1}.}}
This theorem is an immediate consequence of Theorem~\ref{thm:CMR1}
(which can be applied to CMR percolation using any
pair of replicas $\ell,m$) since $D_1^{\ell m}-D_2^{\ell m}
=|Q^{\ell m}|$.

\noindent {\bf {Proof of Theorem~\ref{thm:ultrametric2}.}}
The identity $Q^{12} = D_a -D_d$ which involves only two out of three
replicas may be rewritten in terms of three-replica densities as
\begin{equation}
Q^{12} = D_{aa} +D_{ad} -D_{da} -D_{dd} \, .
\end{equation}
The corresponding formulas for the other overlaps are
\begin{equation}
Q^{13} = D_{aa} -D_{ad} +D_{da} -D_{dd} \, ,
\end{equation}
\begin{equation}
Q^{23} = D_{aa} -D_{ad} -D_{da} +D_{dd} \, .
\end{equation}
Combining these three equations with the identity
$D_{aa} +D_{ad} +D_{da} +D_{dd} = 1$, one may solve for the
$D_{\alpha \alpha'}$'s to obtain
\begin{equation}
D_{aa} = (Q^{12}+Q^{13}+Q^{23}+1)/4 \, ,
\end{equation}
\begin{equation}
D_{ad} = (Q^{12}-Q^{13}-Q^{23}+1)/4 \, ,
\end{equation}
\begin{equation}
D_{da} = (-Q^{12}+Q^{13}-Q^{23}+1)/4 \, ,
\end{equation}
\begin{equation}
D_{dd} = (-Q^{12}-Q^{13}+Q^{23}+1)/4 \, .
\end{equation}
In the equilateral triangle case 
where $|Q^{12}|=|Q^{13}|=|Q^{23}|$
and also $Q^{12}Q^{13}Q^{23}>0$
(from Property P5$'$),
one sees that the corresponding values of the $D_{\alpha,\alpha'}$'s
lie in ${\bf R}^{4,+}_{\mbox {ultra}}$ with 
$x_{(1)}>x_{(2)}=x_{(3)}=x_{(4)}$. In the isosceles triangle
case (with $Q^{12}Q^{13}Q^{23}>0$), one is again in 
${\bf R}^{4,+}_{\mbox {ultra}}$, but this time with 
$x_{(1)}>x_{(2)}>x_{(3)}=x_{(4)}$.

\noindent {\bf {Proof of Theorem~\ref{thm:ultrametric3}.}}
Here we use Lemma~\ref{lemma3} to study the densities (fractions of $N$)
$D^3_{\alpha \alpha',j}$ (where $\alpha$ and $\alpha'$ are a or d)
of the $j$'th largest cluster in $\Lambda_{\alpha \alpha'}$ formed
by the bonds $\{ij\}$ with ${\bf b}_{ij} {\bf n}_{ij}^3 =1$. 
Similarly to the proof of Theorem~\ref{thm:TRFK2}, we have 
\begin{equation}
S_1^{\mbox {RG}}(N \,D_{\alpha \alpha'}, \frac{4 \beta^2}{N}
+ {\mbox {O}} (N^{-3/2})) \, >> \, N\, D_{\alpha \alpha',1}^3
\, >> \, 
S_1^{\mbox {RG}}(N \,D_{\alpha \alpha'}, \frac{4 \beta^2}{N}
+ {\mbox {O}} (N^{-3/2})) \, 
\end{equation}
and use this to conclude that 
$D_{\alpha \alpha',1}^3 - \theta(4 \beta^2D_{\alpha \alpha'})
\, D_{\alpha \alpha'} \to 0$ and $D_{\alpha \alpha',2}^3 \to 0$.
Noting that $D \,  \theta (4 \beta^2 D)$ is an increasing function of $D$,
the rest of the proof follows as for
Theorem~\ref{thm:ultrametric2}.


\bigskip

\noindent {\bf Acknowledgments.}  
The research of CMN and DLS was supported in part by 
NSF grant DMS-06-04869.
The authors thank Dmitry Panchenko, 
as well as Alexey Kuptsov, Michel Talagrand
and Fabio Toninelli for useful discussions
concerning the nonvanishing of the overlap for supercritical
SK models. They also thank Pierluigi Contucci and Cristian Giardin\`{a}
for useful discussions about Property P5 for three overlaps.

	
\newpage
\pagestyle{empty}

\small

\bibliographystyle{unsrt}

\bibliography{refs93007}

\end{document}